\documentclass[dvips,12pt]{emulateapj}
\usepackage[english]{babel}
\usepackage{amsmath}
\usepackage{amssymb}
\usepackage{url}
\usepackage[dvips]{color}
\usepackage{natbib}

\usepackage[dvipdfm,hypertexnames=false]{hyperref} 
\newcommand{\nudot}{\dot{\nu}}
\newcommand{\hzs}{\rm\,Hz\,s^{-1}}
\newcommand{\be}{\begin{equation}}
\newcommand{\ee}{\end{equation}}
 
\newcommand{\beq}{\begin{eqnarray}}
\newcommand{\eeq}{\end{eqnarray}}

\begin{document}

\title{Evidence of Fast Magnetic Field Evolution in an Accreting Millisecond Pulsar}
\author{A. Patruno \altaffilmark{1}} \altaffiltext{1}{Astronomical Institute
  ``Anton Pannekoek,'' University of Amsterdam, Science Park 904, 1098
  XH Amsterdam, Netherlands} 

\begin{abstract}
\noindent

The large majority of neutron stars (NS) in low mass X-ray binaries
(LMXBs) have never shown detectable pulsations despite several decades
of intense monitoring. The reason for this remains an unsolved problem
that hampers our ability to measure the spin frequency of most
accreting NSs. The accreting millisecond X-ray pulsar (AMXP) HETE
J1900.1--2455 is an intermittent pulsar that exhibited pulsations at
about 377 Hz for the first 2 months and then turned in a non-pulsating
source. Understanding why this happened might help to understand why
most LMXBs do not pulsate. We present a 7 year long coherent timing
analysis of data taken with the \textit{Rossi X-ray Timing Explorer}
(\textit{RXTE}). We discover new sporadic pulsations that are
detected on a baseline of about 2.5 years. We find that the pulse
phases anti-correlate with the X-ray flux as previously discovered in
other AMXPs. We place stringent upper limits of 0.05\% rms on the
pulsed fraction when pulsations are not detected and identify an
enigmatic pulse phase drift of $\sim180^{\circ}$ in coincidence with
the first disappearance of pulsations. Thanks to the new pulsations we
measure a long term spin frequency derivative whose strength decays
exponentially with time.  We interpret this phenomenon as evidence of
magnetic field burial.

\end{abstract}
 \maketitle

\section{Introduction}

Some neutron stars in LMXBs have magnetic fields which are
sufficiently strong to truncate the accretion disk and channel plasma
along the field lines. The NS rotation modulates the X-ray emission
emerging from the hot spots plus the accretion shocks that form close
to the NS surface. Detecting their spin has several
important implications for understanding how millisecond radio pulsars
are recycled, how the magnetosphere and the accretion disk interact
and whether sub-millisecond NSs can be formed.

Accreting millisecond X-ray pulsars in LMXBs spin with periods of less
than 10 ms and are powered by channeled accretion. Only a small
fraction of neutron stars in LMXBs are AMXPs (see \citealt{pat10b} for
the AMXP list and \citealt{pap11a} for the most recent system
discovered), the largest majority not showing accretion powered
pulsations with a fractional rms amplitude (see Eq. 4
in~\citealt{har08} for a definition) smaller than $\sim1\%$
(\citealt{vau94, dib05, pat10e}). The reason for this is still a
puzzle and several models attempt to provide an explanation by
invoking gravitational lensing\citep{oze09}, pulse smearing in a hot
electron cloud \citep{tit02}, rotation and magnetic pole alignment
\citep{lam09,lam09b}, MHD instabilities \citep{rom08} and burial of
the magnetic field \citep{cum01}.

This paradigm remained almost unchanged until 2007-2008, when three
AMXPs showed intermittent pulsations: HETE J1900.1-2455
\citep{kaa06,gal07}, Aql X-1 \citep{cas08} and SAX J1749.8-2021
\citep{gav07, alt08, pat09}. In these systems X-ray pulses appear and
disappear on timescales that range from a few hundred
seconds up to several days. In this respect the discovery of
intermittent AMXPs has been an important breakthrough since it was
suggested that most of the non-pulsating LMXBs might indeed
sporadically pulsate.

Particularly interesting is the behavior of HETE J1900.1--2455 which
showed persistent pulsations for the first 22 days until the
occurrence of a flare in the lightcurve on July 8, 2005 (MJD
53,559). Right after this event the pulsations disappeared and
reappeared at different intervals with the source now becoming
intermittent until August 20, 2005 (MJD 53,602). After this date
pulsations disappeared and they were only tentatively detected
with fractional amplitude of 0.29\% rms, by summing the power spectra
of 137 ks of data \citep{gal08}. This is so far the smallest
fractional amplitude ever reported for an AMXP. This slow transition
from a normal AMXP to a non-pulsating LMXB might therefore be the key
to understand why most LMXB do not pulsate.

The outburst of HETE J1900.1--2455 has lasted for $\sim7$ years since
the first discovery on June 15, 2005 and is still ongoing at the
moment of writing this Letter. We present a coherent timing
analysis of the 7 years of data collected. We identify an enigmatic
$180^{\circ}$ drift in the pulse phases discovery during the July 
flare, and we identify new pulsations in a few data segments that
extend to a few years after the last robust detection on August
20, 2005. We use these new pulsations to measure the behavior of the
spin frequency derivative over a baseline of 2.5 years. We then
discuss how these findings might help to understand the pulse
formation mechanism and the behavior of non-pulsating LMXBs.

\section{X-ray Observations and Data Analysis}\label{sec:2}

 We used all high time resolution data taken during the lifetime of
 the \textit{RXTE} with the Proportional Counter Array. We used
 data modes with time resolution of
 $2^{-20}$ s (GoodXenon) and $2^{-13}$ s (Events-122$\mu\,s$).  We
 selected an energy band which spans approximately the range 2--16 keV
 (absolute channels 5--37) to maximize the signal-to-noise ratio (S/N)
 of the pulsations. The data are barycentered using the JPL DE405
 ephemeris at the best determined optical position of HETE
 J1900.1-2455 \citep{fox05} and are cleaned according to standard
 procedures with X-ray bursts removed from our analysis.

 Pulsations are constructed by folding the data in segments of length
 $300$, $500$, $\approx3000$ seconds (i.e., orbit-long \textit{RXTE}
 observations), a few to several hours (i.e., daily \textit{RXTE}
 observation) and very long data stretches that include all data that
 fall within the decoherence timescale (see Section~\ref{sec:3}).  The
 choice of different timescales is made to inspect the presence of
 rapid episodes of high amplitude pulsations (which would be missed in
 long-time averages) or whether pulsations are continuously present
 but with a very low amplitude (by averaging large data stretches).
 The first folding iteration uses the orbital and spin solution
 reported in \citet{kaa06}. We then fit our pulse profiles with a
 sinusoid plus a constant to determine the pulse time of arrivals
 (TOAs) and their pulsed fractions. We set the confidence level for
 the detection of pulsations at 3.6$\sigma$ defined as the ratio
 between the pulse amplitude and its statistical error
 (see~\citealt{pat10c}). This value is
 chosen to guarantee less than one false pulse detection when
 considering the entire amount of trials ($N_{trials}\approx 3000$).
 We then fit the TOAs detected until MJD 53,602 with a Keplerian
 circular orbit and a constant spin frequency $\nu_s$ with the
 software TEMPO2 \citep{hob06,edw06} and repeat the entire folding
 procedure until we reach convergence for our timing solution. The
 reason why we fit the TOAs until MJD 53,602 is that the spacing
 between detected pulses is sufficiently dense to avoid over-fitting
 of the orbital parameters (see for example \citealt{har09} for a
 discussion).

\section{Results}\label{sec:3}

The orbital solution of HETE J1900.1--2455 is reported in
Table~\ref{tab1}. We find consistency between our results and those of
\citet{kaa06} within the statistical uncertainties, although our
orbital period has an order of magnitude higher precision thanks to
the longer baseline (\citealt{kaa06} solution refers to data
up to MJD 53,559). The precision $\sigma_{P_{orb}}$ of our orbital
period $P_{orb}$, is such that $N_{orb}\,\sigma_{P_{orb}} < P_{orb}$,
where $N_{orb}\approx 40,000$ is the number of orbits that HETE
J1900.1--2455 completes in 7 years. Therefore we can extend our
orbital solution to the entire baseline spanned by the \textit{RXTE}
observations.

In principle, $\nu_s$ is also known with sufficient precision from
\citet{kaa06} that we can confidently predict the pulse phases at any
given epoch. However, there is a fundamental complication in this case
represented by the poor knowledge of the spin frequency derivative
$\nudot_s$. The presence of a spin frequency derivative has a
particularly dominant effect in HETE J1900.1--2455 given its long
observational baseline.  We also need to consider the presence of
timing noise and the systematic errors associated with the X-ray
position of the source as given by \textit{Chandra}
\citet{fox05}. This introduces a spurious pulse frequency derivative
$\nu_p$ with magnitude of approximately $-10^{-14}\rm\,Hz\,s^{-1}$
that needs to be taken into account. The variability of $\nu_p$
induces decoherence of the signal which is proportional to the
strength of the pulse frequency derivative $\nudot_p$:
\begin{equation}
\tau_{decoh} = \sqrt{\frac{1}{\nudot_p}}.
\end{equation}
This value gives a maximum baseline of $\approx 40$ d and $\approx
120$ d for $\nudot_p=10^{-13}\hzs$ and $\nudot_p=10^{-14}\hzs$,
respectively.  Therefore we cannot fold very long data stretches
without smearing the pulsations (if present).  Furthermore, even
assuming that $\nudot_s$ is zero and no timing noise is present in the
data, we are limited by the decoherence time introduced by the
spurious pulse frequency derivative due to the source positional
error, which is of the order of 100 days.

\begin{deluxetable}{@{} lr @{}}
\tabletypesize{\footnotesize}
\tablecolumns{2}
\tablewidth{0pt}
\tablecaption{Orbital Solution of HETE J1900.1--2455 
  \label{tab1}}
\startdata
\hline\hline\\[-1.5 ex]
Orbital period, $P_{\rm orb}$ (s)&
  4995.2630(5)\\
Orbital period derivative, $\dot{P}_{\rm orb}$ ($10^{-10}$~s~s$^{-1}$) \tablenotemark{a} &
  $<$1.2\\
Projected semi-major axis, $a_{\rm x} \sin i$ (light-ms) &
  18.44(2)\\
Time of ascending node, $T_{\rm asc}$ (MJD, TDB) &
  53549.130943(9)\\
Eccentricity, $e$ ($95\%$ confidence level) &
  $< 4\times 10^{-3}$
\enddata
\end{deluxetable}

\subsection{New Pulse Episodes}

We detect 4 new pulse episodes after MJD 53,602 with an amplitude
between 1\% and 0.5\% rms and one between MJD 53,584 and MJD 53,596
with an amplitude of 0.3\% rms.  We also find two marginal detections
between MJD 54,865 and 54,967 ($3\sigma$) and between 55,081 and
55,170 ($3.2\sigma$).  The two marginal detections have fractional
amplitudes of 0.1\% and 0.15\% rms, respectively. However, we do not
include them in our analysis since they need confirmation. Robust
detections are made in segments of different length, with higher
amplitudes found in sporadic and short data segments whereas the
lowest amplitudes are found in observations-long data segments. The
new pulse episodes are not detected in coincidence or close to bursts
observed by \textit{RXTE}, with a minimum time interval of a few days
between the pulse episode and its closest burst. The first pulse
episode after MJD 53,602 is detected at MJD 53,624, whereas the last
one detected appears at MJD 54,499 giving a
baseline of more than 2.5 years to perform coherent timing
studies. Upper limits on the pulsed fractions range from 1\% rms down
to 0.05\% rms (95\% confidence level) with most of the segments having
upper limits of about 0.5\% rms.

\subsection{Spin Frequency Derivative}
\begin{figure*}[t!]
  \begin{center}
    \rotatebox{0}{\includegraphics[width=1.0\textwidth]{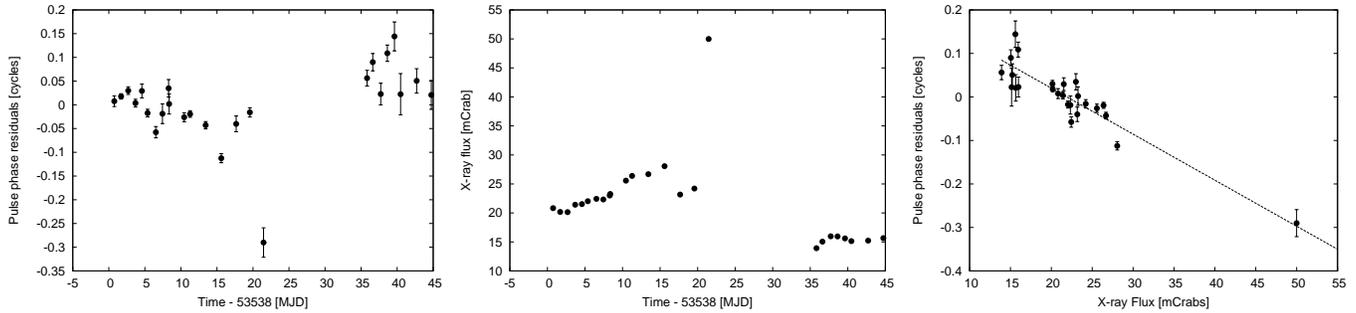}}
  \end{center}
  \caption{\textbf{Left Panel:} Phase residuals of HETE J1900.1--2455
    calculated up to MJD 53,582 when pulsations are sufficiently
    densely sampled. Each point is an observation long average. The
    residuals are calculated with the correlation coherent technique
    (see main text). \textbf{Central Panel:} 2-16 keV X-ray lightcurve of
    the same data. Note that the plot shows only data in which
    pulsations are detected. \textbf{Right Panel:} Anti-correlation between the
    pulse phase residuals and the X-ray flux.}\label{fig1}
\end{figure*}
All AMXPs discovered so far show erratic variations of the pulse
phases on timescales of a few hundred of seconds up to several months
that seem at first sight unrelated with true spin variations of the
NS. These phase variations, commonly called X-ray
timing noise, have been found to be correlated with variations of the
X-ray flux in at least six AMXPs \citep{pat09f}.

To verify whether such a correlation is also present in HETE
J1900.1--2455 we use the correlation coherent analysis described in
\citet{pat10b}. With this method we fit a pulse frequency and frequency
derivative to the TOAs and then choose the two parameters that
minimizes the $\chi^2$ of the linear fit between phase and flux. This
is different than choosing the $\nu_s$ and $\nudot_s$ that minimize the
pulse phase residuals as it is usually done in standard coherent
timing analysis. We find that there is a clear anti-correlation
between the pulse phase and the X-ray flux. The data points exhibit a
tight anti-correlation when considering data points up to MJD 53,582
(see Figure~\ref{fig1}).  The $\chi^2$ of the fit although still
unacceptable ($\chi^2/dof \approx 4$, with 23 dof) is
much better than what can be obtained with standard coherent analysis
($\chi^2/dof \approx 11$ with 23 dof). When considering the whole data
collection (including the sparse detections up to MJD 54,499) the
anti-correlation is still present but with larger scattering and a
worse overall $\chi^2$ (but still significantly better than
what can be obtained with standard coherent analysis).

The reason why the anti-correlation becomes worse after MJD 53,582
might indicate that $\nudot_s$ is not constant as we are assuming in our
 fit. We therefore split the data in seven overlapping
intervals of different length and measure the $\nu_s$ and $\nudot_s$ in
each segment with the correlation coherent analysis method.  We are
forced to use overlapping intervals because the data quality is not
sufficient to allow a measurement of $\nudot_s$ for independent non-overlapping
intervals. Before fitting each interval we change the reference epoch
of our ephemeris so that each $\nu_s$ and $\nudot_s$ refers to a different
epoch that is representative of the interval we are fitting.
The errors are calculated by multiplying by $\sqrt{\chi^2/dof}$ the
statistical errors corresponding to the 90\% confidence interval.

\begin{figure}[t]
  \begin{center}
    \rotatebox{-90}{\includegraphics[width=0.7\columnwidth]{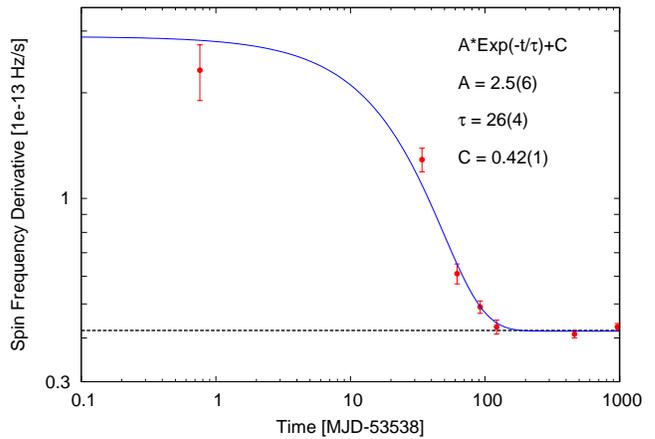}}
  \end{center}
  \caption{Evolution of the spin frequency derivative with time. The
    spin frequency behavior is consistent with an exponential decay
    with an e-folding time $\tau=26\pm4$ days and a base level
    $C=4.2\times10^{-13}\hzs$.}\label{fig2}
\end{figure}

We find that $\nudot_s$ is clearly changing over time and that it is
well described ($\chi^2/dof = 5.9/4$) by fitting an exponential decay
law with a constant level baseline:
\begin{equation}
\nudot_s\left(t\right) = A\,Exp\left(-t/\tau\right) + C
\end{equation}
with an e-folding time $\tau = 26\pm4$ days and a constant level $C$
corresponding to a $\nudot_s = 4.2\times10^{-13}\hzs$.  The behavior of
$\nu_s\left(t\right)$ is well described by the integral of the spin
frequency derivative:
\begin{equation}
\nu_s\left(t\right) = \int \nudot_s\left(t\right) dt = -A \tau Exp\left(-t/\tau\right) + C t + K
\end{equation}
where $K$ is a constant of integration that represents $\nu_s$ at the
beginning of the outburst. The results for $\nudot_s$ are shown in
Figure~\ref{fig2}.  Given that $\nudot_s$ changes over time we re-fold
our data and refit the TOAs considering the different strength
of $\nudot_s$ in different intervals but we find no additional pulse episodes. 

 \subsection{Enigmatic Pulse Phase Drift}

\begin{figure}[t]
  \begin{center}
    \rotatebox{-90}{\includegraphics[width=0.7\columnwidth]{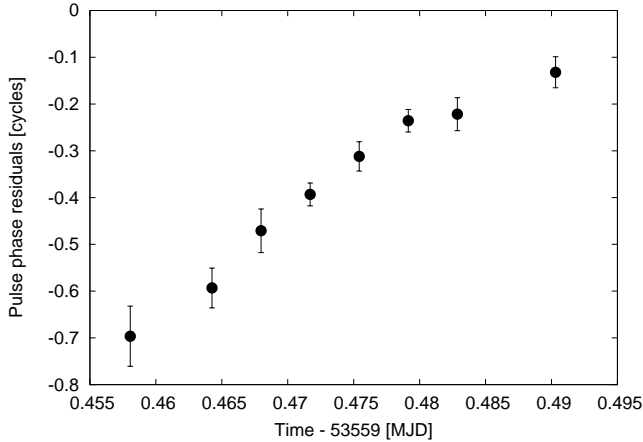}}
  \end{center}
  \caption{Pulse phase drift observed at MJD 53,559. Each data point
corresponds to folded lightcurve segments of 300-500 seconds for a total
observation time of about 3000 seconds. Negative/positive values correspond
to pulsations arriving earlier/later than predicted by the timing model. }\label{fig3}
\end{figure}

 A very interesting feature is evident in the timing residuals in
 coincidence with the flare at MJD 53,559. In observation-long pulse
 profile averages, the pulse phase appears offset by about 0.3 cycles
 with respect to the phases before the flare (see left panel of
 Figure~\ref{fig1}). After the flare the pulsations disappear for
 several days and reappear aligned with the phases before the
 flare. When inspecting 300s to 500s-long pulse profile averages, the
 pulse phases during the flare show a very fast evolution. The phase
 starts 0.7 cycles ($250^{\circ}$) offset with respect to the average
 pulse phase measured in the previous observation (which corresponds
 to phase 0 in Figure~\ref{fig3}) and then drifts back by about
 0.5-0.6 cycles ($\sim180^{\circ}$) towards the phases of the
 pre-flare observations. In Figure~\ref{fig3} a negative/positive
 phase residual means that pulsations arrive earlier/later than
 predicted by the model. An offset of -0.7 cycles is equivalent to an
 offset of +0.3 cycles, but the interpretation with the negative sign
 is the correct one because the pulsations show a linear drift with
 each successive pulsation lagging the previous one. The timescale for
 the drift is $\approx3000$ s and pulsations have fractional rms
 amplitude which remains approximately constant within the statistical
 errors with a slight excess in the middle of the observation. There,
 the fractional amplitude reaches 1.6\% rms and stays
 at 1.1\% rms in the rest of the observation. The pulsations disappear
 after this event until MJD 53,573 with 95\% c.l. upper limits
 between 0.35\% rms and 1.5\% rms. The relatively larger error-bar of
 the pulse phase obtained by folding the entire data record at the
 high flux point (as shown in Figure 1) is therefore artificial
 because the pulse amplitudes are partially smeared out by
  the fast pulse phase drift.

\section{Discussion}

The detection of sporadic pulse episodes in the 2.5 years following
the last robust detection at MJD 53,602 strengthens the suggestions of
\citet{gal08} that pulsations might be always present in HETE
J1900.1--2455. The pulses we detect might represent the ``tip of the
iceberg'' of very weak pulsations that are present at a level of
$\lesssim0.1\%$ rms, since our upper limits reach the most stringent
value of about 0.05-0.5\% rms. This suggestion is reinforced by our
detection of pulses with amplitudes as low as 0.3\% rms. Such
low values can be reached only because we have an initial timing
solution for the orbit and the NS spin which is sufficiently
precise to allow a coherent analysis over the entire 7 years of
observations. It is not possible, with current instrumentation, to
inspect any other non-pulsating LMXB with a comparable X-ray flux and
place such extreme upper limits for the pulsed fractions with
incoherent timing techniques. Existing upper limits on the pulse
fractional amplitudes in non-pulsating LMXBs are usually of the order
of 1\% rms.
\begin{figure}[t!]
  \begin{center}
    \rotatebox{-90}{\includegraphics[width=0.7\columnwidth]{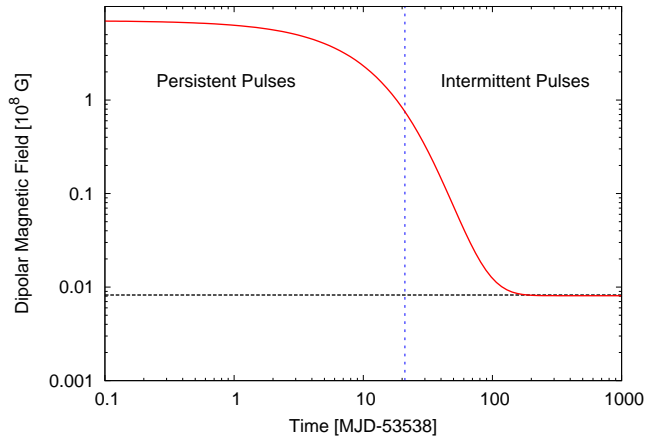}}
  \end{center}
  \caption{Evolution of the magnetic field according to
    Eq.(\ref{spinup}) for
    $\dot{M}_{-10}=10^{-9}\rm\,M_{\odot}\,yr^{-1}$ and an
    exponentially decaying $\nudot_s$ as shown in Figure~\ref{fig2}. The vertical 
dashed blue line represents the time when pulsations first disappear and corresponds to a field of $7\times10^{7}$ G.}\label{fig4}
\end{figure}
The baseline available for measuring the spin frequency and the
spin-up is $\sim2.5$ years and this is an unprecedented possibility
for AMXPs, since accretion torques have been investigated so far only
on relatively short timescales reaching at most $\sim100$
days~\citep{pat10c}. The spin up of HETE J1900.1--2455 follows an
exponential decay with an e-folding timescale of $26\pm4$ days and a
constant baseline of about $4\times10^{-13}\hzs$.  The spin frequency
derivative expected from accretion torques that develop at the
disk-magnetosphere boundary of standard thin disks truncated by a
constant dipolar magnetic field is:
\begin{equation}
\nudot_s \approx 2.3\times 10^{-14}\xi^{1/2}\dot{M}_{-10}^{6/7}B^{2/7}_8 M_{1.4}^{-5/21}R_{10}^{6/7} \hzs. \label{spinup}
\end{equation}
where $\dot{M}_{-10}$ is the mass accretion rate in units of $10^{-10}
M_\odot/\mbox{yr}$, $B_{8}$ the dipolar magnetic field at the poles in
units of $10^8$ G, $M_{1.4}$ the mass of the NS in units of $1.4
M_{\odot}$, $R_{10}$ its radius in units of $10$ km. The parameter $\xi$ is 
introduced to account for the uncertainties
in the torque at the edge of the accretion disc and is 
in the range $\xi\approx 0.3-1$ \citep{psa99}.

As noticed by \citet{gal08}, the lightcurve of HETE J1900.1--2455
shows an erratic behavior with large variations in X-ray flux on
timescales of tens of days. The flux averaged over several tens of
days is, however, rather constant (we find variations of at most 20\%
in X-ray flux between the averages of our seven intervals) and is
certainly not exhibiting an exponential drop. Therefore the
exponential decay of $\nudot_s$ cannot be related solely with variations
in $\dot{M}$ (if we assume that $L_{X}\propto\dot{M}$; see, however,
~\citealt{van01} for a criticism), neither with variations of $M$ and
$R$ which stay basically constant throughout the outburst. If the
magnetic field is responsible for the variation of $\nudot_s$, then the
$B$ field has to decay approximately exponentially with a very short
e-folding time of $\sim10$ days.

Magnetic field burial models predict an exponential decay of the
external magnetic field which is screened by freshly accreted plasma
\citep{cum01, cum08}. As matter is accreted to the polar caps it will
eventually spread laterally and bury the field underneath. If the new
material accumulates on the NS surface on a timescale which
is much shorter than the Ohmic diffusion timescale, the magnetic field
is (partially) screened. In these models the field suppression
operates on timescales that vary significantly and depends on several
assumptions.  For example, \citet{cho02} have shown that screening
timescales of $\sim1$ yr can be achieved provided that the no magnetic
buoyancy is present (see also \citealt{pay04, pay07, wet10}). The fact
that our fit requires a constant level can also be naturally explained
in this scenario by considering that the magnetic field is not
efficiently screened once its value is so low that channeled accretion
becomes difficult (see \citealt{kon04} and compare our
Figure~\ref{fig2} with their Figure~10; \citealt{zha06}). Once the
outburst is over, the magnetic field can re-emerge on the Ohmic
diffusion timescale \citep{cum01} and return to its initial value of
$\sim10^8$ G. For illustrative purposes we plot in Figure~\ref{fig4}
the $B$ field at the NS poles of HETE J1900.1--2455 where we
use Eq.(\ref{spinup}) and assume a constant $\dot{M}_{-10}=10$,
$\xi=1$, $M_{1.4}=1$ and $R_{10}=1$. The AMXP has a magnetic field at
the beginning of the outburst of $B_i\sim5\times10^{8}$ G and a final
field of $B_f\sim7\times10^5$ G, which is significantly less than the
minimum field necessary to truncate the accretion disk \citep{psa99}.
This could in principle be related to the extraordinary low rms
amplitudes of pulsations in HETE J1900.1--2455.

A similar behavior might not be observed in other \textit{persistent}
AMXPs possibly because of the substantially smaller mass accretion
rate which is one to two orders of magnitude lower than in HETE
J1900.1--2455. If the timescale for the magnetic screening scales
inversely with the mass accretion rate \citep{kon04} then the magnetic
field of persistent AMXPs might require one to several years to
substantially decrease. Since the outburst duration of persistent
AMXPs is at most 100 days, the screening mechanism cannot reduce the
strength of the magnetic field below the level necessary to channel
plasma along the field lines.

If magnetic screening is the correct interpretation of our findings,
then the anti-correlation between phase and X-ray flux might possibly
also be explained with variations of the magnetosphere (and thus on
the position of the polar caps) that responds to variations in the
amount of accreted material (see also theoretical investigations of
this problem in~\citealt{lon12, kaj11, pou09, lam09}). However, the
anti-correlation does not work on very short timescales of the order
of a few hundred seconds. In particular, the sudden
$\approx180^{\circ}$ (0.5 cycles) pulse phase drift observed at MJD
53,559 remains an enigmatic event and it might contain the key to
understand why pulsations disappeared for the first time right after
this event.

\acknowledgements{I acknowledge support from the Netherlands
  Organization for Scientific Research (NWO) Veni fellowship. I would
  like to thank B. Haskell, T. Tauris and J. Braithwaite for
  interesting discussions.}

\end{document}